\documentclass{optica-article}

\journal{opticajournal} 
\articletype{Preprint}

\begin{document}

\title{Caustic analysis of partially coherent self-accelerating beams: Investigating self-healing property}
\noindent
 \textbf{Peiyu Zhang,\authormark{1} Kaijian Chen,\authormark{1} Chuanhou Zhang,\authormark{1} Jiafang Liang,\authormark{1} Shengyu Deng,\authormark{1} Peilong Hong,\authormark{2,3,*} Bingsuo Zou,\authormark{4} and Yi Liang\authormark{1,3,5,*}}

\address{\authormark{1}Guangxi Key Lab for Relativistic Astrophysics, Center on Nanoenergy Research, School of Physical Science and Technology, Guangxi University, Nanning, Guangxi 530004, China\\
\authormark{2}School of Optoelectronic Science and Engineering, University of Electronic Science and Technology of China (UESTC), Chengdu 610054, China\\
\authormark{3}The MOE Key Laboratory of Weak-Light Nonlinear Photonics, TEDA Applied Physics Institute and School of Physics, Nankai University, Tianjin 300457, China\\
\authormark{4}School of Physical Science and Technology and School of Resources, Environment and Materials, Key Laboratory of new Processing Technology for Nonferrous Metals and Materials, Guangxi University, Nanning 530004, China\\
\authormark{5}State Key Laboratory of Featured Metal Materials and Life-cycle Safety for Composite Structures, Nanning 530004, China}

\email{\authormark{*}plhong@uestc.edu.cn} 
\email{\authormark{*}liangyi@gxu.edu.cn}

\begin{abstract*}
We employed caustic theory to analyze the propagation dynamics of partially coherent self-accelerating beams such as self-healing of partially coherent Airy beams. Our findings revealed that as the spatial coherence decreases, the self-healing ability of beams increases. This result have been demonstrated both in simulation and experiment. This is an innovative application of the caustic theory to the field of partially coherent structured beams, and provides a comprehensive understanding of self-healing property. Our results have significant implications for practical applications of partially coherent beams in fields such as optical communication, encryption, and imaging.
\end{abstract*}



\section{Introduction}
The interference phenomenon is a direct manifestation of a beam's wave nature, and the degree of interference is determined by the beam's coherence \cite{RN1}. Coherence is an intrinsic property of the beam that can be considered a degree of freedom in controlling its behavior and propagation property. Understanding the theory of partially coherent beams is essential in comprehending the concept of structured beams, which result from the interference and superposition of beams \cite{RN2}. Moreover, by controlling the spatial coherence structure of a beam, coherence can provide robustness to complex media and atmospheric turbulence \cite{RN3,RN4,RN5}. This capability is particularly useful in optical particle trapping \cite{RN6,RN7,RN8,RN9}, high-resolution imaging \cite{RN10,RN11,RN12}, and optical communications \cite{RN13}, where structured light fields have brought significant changes to these fields. 

Usually, the coherence of a modulated structured beam has a significant effect on its propagation properties \cite{RN14}. For example, the self-accelerating beam, which is the most typical representative of coherent structured light fields, possesses the characteristics of non-diffraction, self-acceleration, and self-healing \cite{RN15}. Since its discovery, extensive researches have been conducted on the generation, propagation, and applications of the self-accelerating beam \cite{RN16,RN17,RN18,RN19}. It is found that, when the spatial coherence of self-accelerating beams such decreases, the intensity of the main lobe of the beam decays more rapidly during free-space propagation, resulting in a shorter non-diffracting distance and reduced lateral shift during propagation \cite{RN20}. However, numerical simulations have demonstrated that partially coherent self-accelerating beams have a stronger self-healing ability than conventional fully coherent accelerating beams \cite{RN21}. In addition, several theories have been proposed to explore the physical mechanisms of the self-healing property of coherent structured beams, including transverse power flow analysis based on the Babinet principle \cite{RN24}, non-diffraction explanation \cite{RN25,RN26}, caustic theory \cite{RN27,RN28,RN29}, and wave theory \cite{RN30,RN31}. While these theories have provided powerful explanations for the self-healing properties of coherent beam, the self-healing of partially coherent structured beam remains lacking a comprehensive theoretical framework and corresponding experimental demonstrations.

In essence, a caustic is the envelope of rays that are reflected or refracted by curved surfaces or objects. The theory of caustic has been extensively studied in the context of self-healing, and it has been shown to provide valuable insights into the behavior of self-accelerating beams \cite{RN28,RN33}. One of the advantages of the caustic model is its simplicity. By considering only the geometric properties of rays and caustics, it allows us to predict the propagation behavior of beams without having to resort to more complex models. This makes it a valuable tool for both theoretical and practical applications, from designing optical beams to analyzing the behavior of light in complex environments.

In this study, we proposed partially coherent self-accelerating beams based on the caustic theory and then used the caustic theory to simulate the propagation dynamics of self-accelerating beams such as Airy beams during their self-healing process. We provided an explanation for the enhanced self-healing ability as spatial coherence decreases. Subsequently, we validated our proposed caustic theory of partially coherent beam through simulations and experiments. This fills a gap in the field of partially coherent structured beam in the context of caustic theory. Our findings have significant implications for the use of partially spatially coherent beam in optical micromanipulation, optical imaging, and optical communication in complex environments. By shedding light on the self-healing properties of partially coherent structured beam, our study opens up new avenues for the development of advanced optical technologies.

\section{Theory}

To analyze the geometric properties of the caustic, the phase of the jump integral (also known as the angular spectrum) is utilized. Consider a  self-accelerating beam propagating in the $z$ direction in free space, which is partially obstructed by an obstacle located along the transverse direction $x$. The starting point for analysis is the full phase of the beam's angular spectrum, with power $n$ and symmetry $q$ \cite{RN34,RN35}. Enlightened by previous studies \cite{RN36,RN37}, based on the coherent mode superposition method, we introduce random additional dimensionless spatial frequencies to the phase of the jump integral at $z=0$ to form independent coherent modes. The phase of the independent coherent mode can be expressed as:
\begin{equation}
{\Phi^{(q)}_{n,m}(K;s,\xi)=C_nsgn^q(K'_m) \left | K'_m\right |^n+\xi \left [ k^2-\frac{(K'_m)^2}{2}\right ]+s K'_m}
\label{eq:refname1}
\end{equation}
where $K'_m=K+\Delta K_m$. $K$ is a dimensionless spatial frequency and is the conjugate variable of the normalized transverse spatial coordinate $s=\frac{x}{x_0}$. $\Delta K_m$ comes from the random phase and is related to the coherence length. $x_0$ is a transverse scale characterizing the beam. The normalized propagation coordinate is $\xi=\frac{z\lambda}{2\pi x^2_0}$ , and $k=(\frac{2\pi}{\lambda})x_0$ is the normalized free-space wave number with $\lambda$ being the free-space wavelength. The normalization constant $C_n$ guarantees the same range of phase variation for a given range of $K$. $sgn(·)$ is the sign function and $\left | ·\right |$ is the absolute value. Usually, partially coherent self-accelerating beams can be superposed by coherent modes \cite{RN36,RN37}. Therefore, the coherent function of partially coherent self-accelerating beam can be expressed as:
\begin{equation}\label{eq:refname2}
\begin{split}
& W\left({s}_{1},{s}_{2},\xi\right)=\sum_m  E_m(s_1,\xi)E_m(s_2,\xi)  \\
& E_m(s,\xi)= \int_{-\infty }^{+\infty} \exp\left [ i\Phi^{(q)}_{n,m}(K;s,\xi)\right ]{\mathrm{d}K}
\end{split}
\end{equation}
 Thus, the spatial coherence of the beam is $\mu({s}_{1},{s}_{2},\xi)=W\left({s}_{1},{s}_{2}, \xi\right)/\sqrt{I({s}_{1},\xi)I({s}_{2},\xi)}$. Then, the coherence length $\delta$ can be defined as the transversal distance $\Delta s={s}_{1}-{s}_{2}$ at which the spatial coherence function $\mu({s}_{1},{s}_{2}, \xi)$ decreases to a certain fraction (e.g., $1/\text{e}$ or $1/{\text{e}}^{2}$) of its maximum value.

The ray pattern associated to the beam, and thereby its caustic structure, can be obtained by differentiating the generating function with respect to the internal variable and setting the derivative to zero \cite{RN38}. The caustic structure for each coherent mode is:
\begin{equation}
{\frac{\partial \Phi^{(q)}_{n,m}}{\partial K}=nC_nsgn^{q+1}(K'_m)\left |K'_m\right |^{n-1}-\xi K'_m+s=0}
\label{eq:refname3}
\end{equation}
By computing ray modes with different slopes through Eq. (\ref{eq:refname3}), also known as ray clusters, we can represent the optical field using these ray clusters. As we know, all rays are tangent to a curve which is essentially the envelope formed by the ray clusters, obtained by setting the second partial derivative of ${{\Phi^{(q)}_{n,m}}}$ to zero:
\begin{equation}
{\frac{\partial^2 \Phi^{(q)}_{n,m}}{\partial K^2}=n(n-1)C_nsgn^q(K'_m)\left |K'_m\right |^{n-2}-\xi=0}
\label{eq:refname4}
\end{equation}
This curve is called the caustic, and because the first and second derivatives of the phase function are both zero on the caustic, it can be ensured that the intensity of the light near the caustic has a maximum value.  
One can obtain the equation of the caustic curve from Eq. (\ref{eq:refname4})
 by substituting $K$ into Eq. (\ref{eq:refname3})
  \cite{RN34,RN35}. 

\begin{figure}[ht!]
\centering
\includegraphics[scale=1]{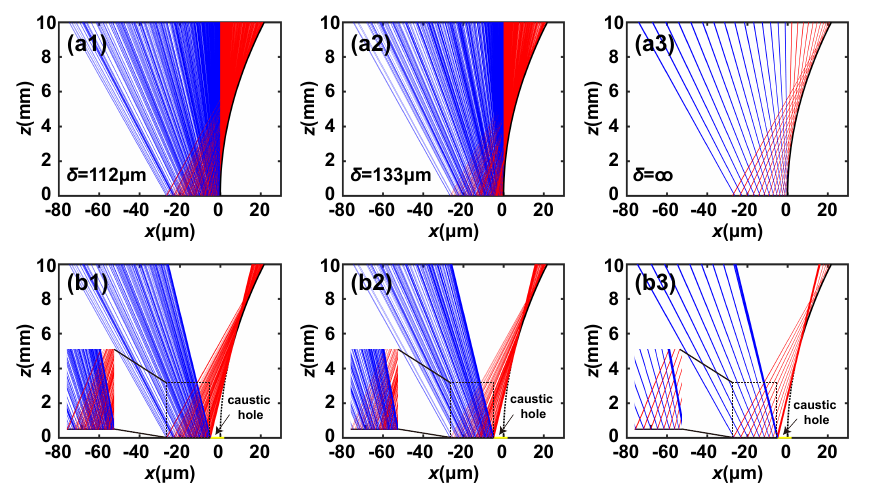}
\caption{Caustic patterns of PCABs with different coherence lengths: 112 $\upmu$m, 133 $\upmu$m, and $\infty$. (a1-a3) Caustic patterns without obstacles. (b1-b3) Caustic patterns with obstacles.}
\label{Fig.1}
\end{figure}

To prove that the propagation dynamics of partially coherent self-accelerating beam can be analyzed successfully by the above theory, here, we take $n=3$, $q=1$, $C_n=1$, $x_0$=14$\upmu$m, $\Delta K_m<K$ and $K\in(-0.4,0.4)$ as an example, generating the caustic patterns of partially coherent Airy beams (PCABs) with coherent lengths of $\infty$, 133 $\upmu$m and 112 $\upmu$m. To make the propagation properties of PCABs with different coherence more evident, we superimpose 50 coherent modes via Eq. (\ref{eq:refname2}), with a random phase added to each coherent mode for PCABs and a constant difference phase for coherent Airy beam. In addition, we change the vertical axis of the above formula from the normalized propagation distance to the propagation distance, and take only the positive half axis, the horizontal axis was transformed from the normalized horizontal space coordinate to the horizontal space coordinate. All the caustic results are shown in Fig.
\ref{Fig.1}.

The speckle pattern depicted in Fig. \ref{Fig.1}(a) exhibits red and blue rays, representing positive and negative spatial frequencies of PCABs, respectively. PCABs can be understood as the superposition of plane waves with different spatial frequencies. The positive spatial frequencies correspond to the forward propagation, which is associated with the main lobe of the beam. It is worth noting that the coherence, phase, and spatial frequency of the beam are interconnected. In a completely coherent beam, the phase relationship remains constant, resulting in spatial frequency components that are evenly distributed in space and exhibit a regular spatial frequency characteristic [Fig. \ref{Fig.1}(a3)]. Conversely, partially coherent beams display phase fluctuations, leading to significant fluctuations and chaotic spatial frequency characteristics [Figs. \ref{Fig.1}(a1, a2)]. This effect becomes more pronounced as the coherence decreases. The rich and random spatial frequencies induce rapid phase changes in the beam, resulting in alterations in the propagation direction, beam spreading, and blurring of its organized structure. As a consequence, PCABs lose their fine structural characteristics. Furthermore, the inclusion of dimensionless spatial frequencies has minimal impact on the caustic curves, yielding approximately identical caustic curves for PCABs with different coherent lengths. These propagation properties of PCABs have been previously demonstrated in Ref. \cite{RN36}.

When an obstacle obstructs the main lobe of PCABs (depicted by the yellow ray), it interferes with a subset of spatial frequencies, causing the PCABs to lose part of their parabolic envelope and form a speckle aperture [Fig. \ref{Fig.1}(b)]. The size of this aperture is determined by the two blocking marginal rays (represented by thick red and blue rays). Importantly, the obstruction of specific rays by an obstacle disrupts a section of the caustic curve. This caustic rupture extends within the spatial region defined by the rays touching the edges of the obstacle, which we refer to as the "blocking limit rays".

As the red rays propagate, the self-healing process of PCABs is initiated. We propose that the self-healing properties of the beam can be attributed to the reconstruction of spatial frequencies. Due to the phase randomness inherent in partially coherent beams, different spatial frequency components have the opportunity to compensate and adjust each other through secondary randomization of the phase [Figs. \ref{Fig.1}(b1, b2)]. This leads to the reshaping of the original spatial frequencies and the restoration of the partially coherent beam's shape. Such a mechanism represents the self-healing behavior of the partially coherent beam. In contrast, completely coherent beams with fixed phases and regular spatial frequency distributions have limited capacity to reshape spatial frequencies when the phase changes [Fig. \ref{Fig.1}(b3)]. As a result, the self-healing ability of completely coherent beams is inferior to that of partially coherent beams. Although more red rays will cross the blocking marginal rays and contribute to the self-healing process during the propagation of PCABs, low-coherent PCABs may require longer propagation distances to achieve optimal self-healing due to the complexity of their original spatial frequencies. Subsequently, we will demonstrate these predictions regarding the self-healing of PCABs through simulations and experiments.

\section{Simulation and experiment}

In this section, we delve into the self-healing properties of PCABs with varying coherence lengths when the main lobe is obstructed by an obstacle in the source plane. Both simulations and experiments, following a methodology similar to that described in Ref. \cite{RN39}, are conducted to comprehensively analyze the entire self-healing process of PCABs. Figure \ref{Fig.2} and video in Appendix A display the intensity distributions of PCABs with different coherence lengths at various propagation distances, offering insights into the evolution of the self-healing process. 

\begin{figure}[ht!]
\centering
\includegraphics[scale=1.2]{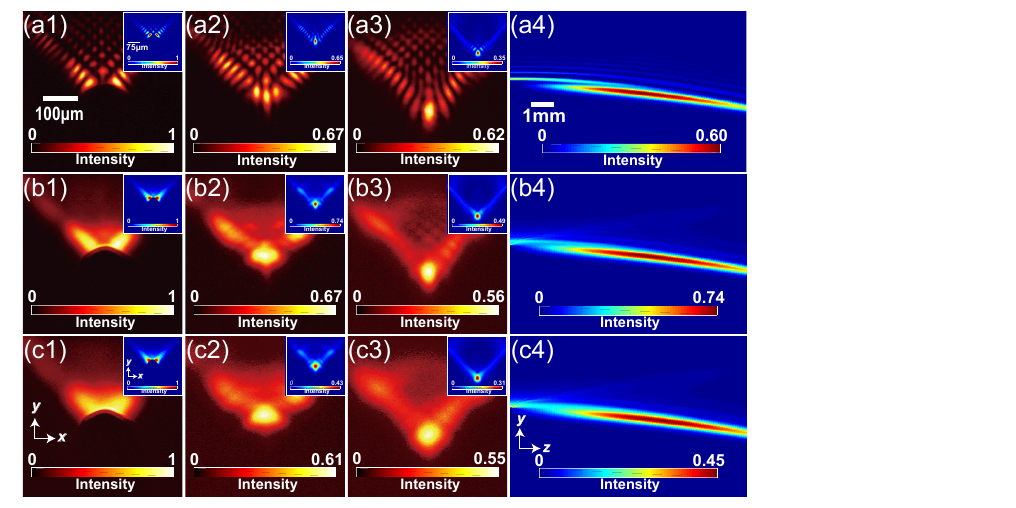}
\caption{Experimental and simulated results of beam profiles of PCABs with different coherences ($\infty$, 133 $\upmu$m, and 112 $\upmu$m) at different propagation positions: (a1-c1)   z= 0 mm. (a2-c2)  z= 5 mm. (a3-c3) z = 10 mm. (a4-c4) The propagation side view of PCABs. The characteristic length $x_0$ and the truncation factor $a$ of PCABs are adopted as 14 $\upmu$m and 0.1.}
\label{Fig.2}
\end{figure}

As shown in Figure \ref{Fig.S1} in the Appendix A, the main lobe of the PCAB becomes larger and the side lobes become smaller as the coherence length decreases. This observation suggests that the finite coherence length acts as a truncation factor for the Airy beam, where reducing the spatial coherence has a similar effect to enlarging the truncation factor. Furthermore, as the coherence length decreases, the fine structure of the Airy beam gradually becomes blurred, and the main and side lobes are no longer clearly defined. In the limit of complete incoherence, the beam profile approximates that of a Gaussian beam.

To experimentally investigate the self-healing properties of PCABs, we blocked the main lobes of PCABs with coherence lengths of $\infty$, 133 $\upmu$m, and 112 $\upmu$m using the same obstacle in the source plane, as shown in Figure \ref{Fig.2}(a1-c1). The corresponding simulated results are displayed in the top right corner for comparison. Both the experimental and simulated intensity distributions have been normalized. Figures \ref{Fig.2}(a2-c2) and \ref{Fig.2}(a3-c3) present the profiles of PCABs with different coherence lengths after propagating distances of 5 mm and 10 mm, respectively. It is observed from both the experimental and simulated results that, despite the obstruction of the main lobe, the PCABs are able to reconstruct themselves, demonstrating their self-healing properties.For a more comprehensive understanding of this process, please refer to the supplemental material, which provides detailed information about the experimental setup and includes a self-healing video.

To quantitatively analyze the impact of coherence on the self-healing properties of PCABs, we defined a factor called intensity profile similarity:
\begin{equation}
{S=\frac{Cov(I_{out},I_{ini})}{Var(I_{out})Var(I_{ini})}}
\label{eq:refname5}
\end{equation}
where $I_{out}$ and $I_{ini}$ represent the final and initial intensity distributions of PCABs with propagation, $Cov(X,Y)$ is the covariance between random distribution $X$ and $Y$, $Var$ is the variance of random distribution $X$. The value of $S$ ranges from 0 to 1, with a higher $S$ value indicating a stronger self-healing ability. 

\begin{figure}[ht!]
\centering
\includegraphics[scale=0.6]{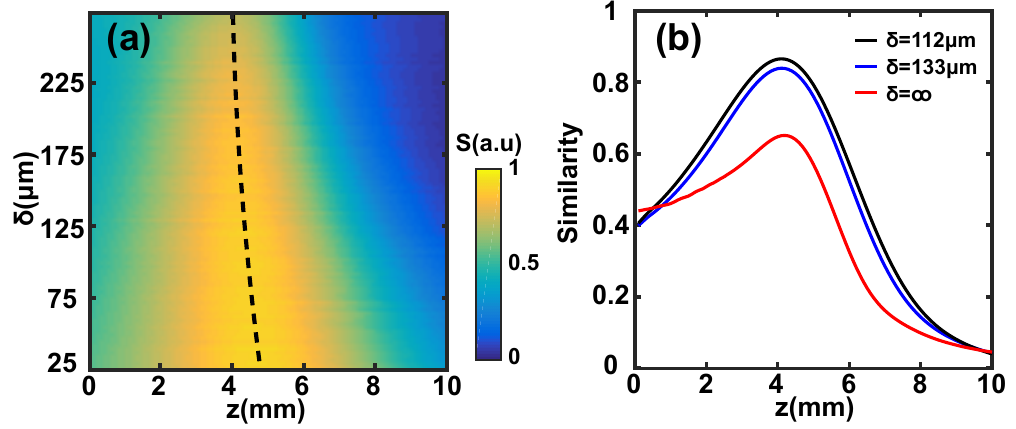}
\caption{(a) The variation of similarity $S$ in the propagation process of PCABs with different coherence lengths $\delta$. The black curve shows the position where the PCABs of different coherent lengths reach the best self-healing state during propagation. (b)  Similarity $S$ of PCABs as a function of propagation distance $z$, for coherence lengths of $\infty$, 133 $\upmu$m, and 112 $\upmu$m. }
\label{Fig.3}
\end{figure}

Thus, we investigated the influence of coherence on the self-healing properties of PCABs using a similarity measure defined in Eq. (\ref{eq:refname5}). We examined how the similarity between the initial PCAB and the propagated beam changes with propagation distance for different coherence lengths. The results, presented in Figure \ref{Fig.3}, reveal the self-healing performance of PCABs at various coherence lengths. As the coherence length decreases, the peak value of the similarity measure $S$, which represents the self-healing ability, increases. This indicates that partially coherent beams exhibit stronger self-healing abilities compared to fully coherent beams. Additionally, based on the experimental data, PCABs with coherence lengths of 112 $\upmu$m, 133 $\upmu$m, and $\infty$ had initial similarity values of 0.75, 0.74, and 0.73, respectively, at $z=0$. As the beams propagated to distances of 5 mm, their similarities increased to 0.85, 0.79, and 0.48, respectively, and then decreased to 0.61, 0.54, and 0.23 at distances of 10 mm. This observation further supports the notion that partially coherent beams possess superior self-healing capabilities. Furthermore, Figure \ref{Fig.3}(a) reveals that PCABs with lower coherence achieve their optimal self-healing state at later propagation distances. All these observations align well with the caustic theory we proposed, which describes the self-healing behavior of partially coherent beams.

\section{Conclusions}
In conclusion, we have successfully applied the caustic theory to analyze the propagation dynamics, including the self-healing phenomenon, of partially coherent self-accelerating beams, specifically focusing on partially coherent Airy beams. Our analysis has revealed that lower coherence levels are associated with improved self-healing capabilities. This comprehensive understanding of the self-healing properties of partially coherent structured beams opens up new avenues for their applications in various fields. By harnessing the enhanced self-healing capabilities, these beams can be utilized for advanced optical manipulation, imaging, and communication systems, among other potential applications.

\begin{backmatter}

\bmsection{Disclosures} The authors declare no conflicts of interest.

\bmsection{Data availability} Other data underlying the results presented in this paper are not publicly available at this time but may be obtained from the authors upon reasonable request.
\begin{figure}[htbp]
\centering
{\includegraphics[scale=0.25]{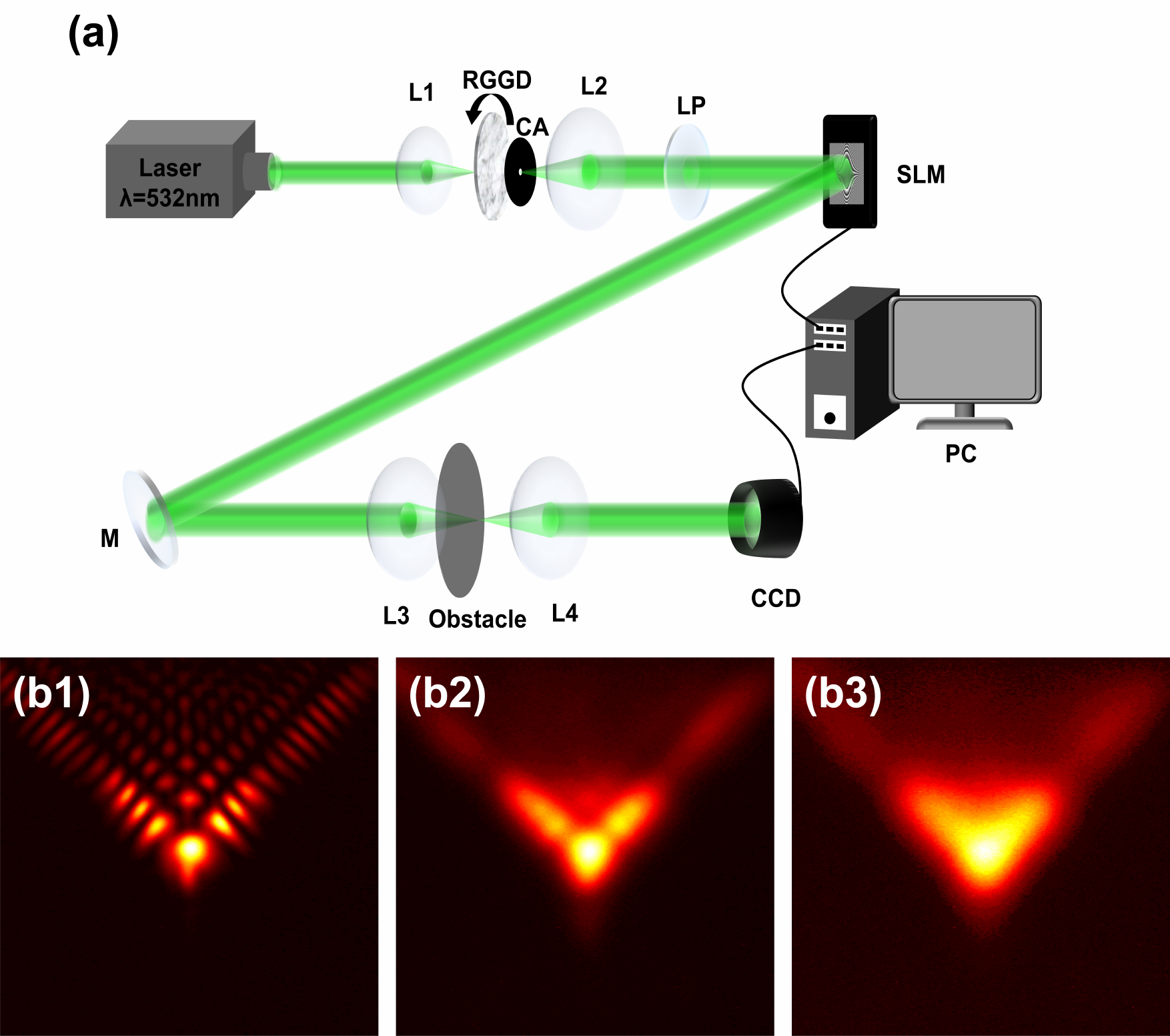}}
\caption{(a) Schematic of experimental setup. L1, L2, L3, L4, thin lens; RGGD, rotating ground-glass disk; CA, circular aperture; M, mirror; LP, linear polarizer; SLM, spatial light modulator; CCD, charge coupled device. (b1-b3) The beam profiles of PCABs with coherence lengths of infinity, 133 $\upmu$m, and 112 $\upmu$m at the source plane. The characteristic length $x_0$ and the truncation factor $a$ of PCABs are adopted as 14 $\upmu$m and 0.1.}
\label{Fig.S1}
\end{figure}

\bmsection{Appendix A} Appendix A includes the experimental setup for the generation and propagation of PCABs, the intensity distribution of PCABs at the source plane and the video of PCABs self-healing process during propagation. Fig. \ref{Fig.S1}(a) shows our experimental setup for generating and propagating PCABs. We begin with a coherent beam at a wavelength of 532 nm, which is focused by lens L1 onto a frosted glass surface. The scattered light from the surface is spatially incoherent, with a Gaussian coherence. By adjusting the distance between lens L1 and the surface, we control the size of the spot on the Random Ground Glass Diffuser (RGGD), which in turn controls the coherence length of the PCABs. The beam then passes through a circular aperture and is collimated by lens L2, reflected by mirror M, and polarized by a linear polarizer to match the requirements of the spatial light modulator (SLM) for incident light polarization. The SLM applies three phase maps to the beam's wavefront, and the modulated beam is Fourier transformed by lens L3 to generate PCABs at the front focal plane of L3. We monitor the self-healing ability of the PCABs using a CCD, which images the beam's main lobe blocked by an opaque obstacle. Finally, we keep the diameter of the coherent Gaussian beam incident on the SLM constant at 6 mm for all experiments. Fig. \ref{Fig.S1}(b1-b3) exhibit the beam profiles of PCABs with coherence lengths of infinity, 133 $\upmu$m, and 112 $\upmu$m in the source plane. Finally, the video site: https://opticapublishing.figshare.com/s/3991e048b2c496c5e045.

\end{backmatter}

\end{document}